\documentclass[12pt]{iopart}

\jl{1}

\usepackage{graphicx}
\usepackage{mathrsfs}
\usepackage{cite}
\usepackage{iopams}
\usepackage{latexsym}
\usepackage{bm}

\newcommand{\mom}{\Pi}
\newcommand{\meas}{\bm{\kappa}}
\newcommand{\tfrac}{\textstyle\frac}

\begin{document}

\title{Radiation transport in diffractive media}

\author{Mattias Marklund}
%\email{mattias.marklund@physics.umu.se}
\address{Department of Physics, Ume{\aa} University, 
  SE--901 87 Ume{\aa}, Sweden}
  
\begin{abstract}
We consider radiation transport theory applied to non-dispersive but refractive 
media. This setting is used to discuss Minkowski's and Abraham's electromagnetic 
momentum, and to derive conservation equations independent of the choice of momentum definition. Using general relativistic kinetic theory, we derive and discuss a radiation gas energy-momentum conservation equation valid in arbitrary curved spacetime with diffractive media.  
\end{abstract}

%\submitted

%\date{\today}
%\maketitle

%%%%%%%%%%%%%%%%%%%%%%%%%%%
\section{Introduction}
%%%%%%%%%%%%%%%%%%%%%%%%%%%

Radiative transfer is a mature area of research, with important application both in laboratory and astrophysical systems (see, e.g., \cite{Mihalas,Milne} and references therein). Much is thus known about the properties of the relevant radiation transport equations in non-dispersive, non-diffractive media in flat spacetime. However, less is known in the case of dispersive and/or diffractive media, and the addition of spacetime curvature reduces the number of relevant publications further (for a representative but incomplete selection see \cite{Milne,Harris,Pomraning,Enome,Bicak-Hadrava,Anderson-Spiegel,Kichenassamy-Krikorian}). 
Different ways to tackle the problems involved in formulating a general theory of curved spacetime radiation transport in diffractive media exists in the literature. In many cases, the treatment of moments of the transfer equations, and their respective conservation equation, is lacking, as is the discussion of the interpretation of the radiation fluid moments. The latter can be viewed as a nontrivial task, since a clear-cut definition of the radiation momentum density is still under discussion (although claimed otherwise by some authors, see, e.g., \cite{Jackson}). There are exceptions in the literature though.  For example, Anderson \& Spiegel introduces an optical geometry on top of the curved spacetime, by introducing an effective metric which incorporates the refractive index of the medium (similar to Gordon \cite{Gordon}). In this effective geometry, the transfer equation takes the standard form, and a treatment of the fluid moments is thus straightforward, although their interpretation is somewhat obscured by the presence of the effective geometry. Moreover, the proper conservation equations for the fluid moments is not derived. 

As noted by many authors, related to the problem of radiative transfer in diffractive media is concept of photon momentum and the Minkowski--Abraham debate. Although the definition of Abraham is preferred due to a number of theoretical reasons (symmetries \cite{Jackson,Landau-Lifshitz}, derivations from microscopic theory \cite{deGroot} etc., see \cite{Brevik} for an overview), measurements have not given a definite answer \cite{Jones-Richards,Jones-Leslie} (for a review, see \cite{Loudon} and references therein). Due to the problem of separating the electromagnetic from the material degrees of freedom \cite{Nelson,Loudon-Allen-Nelson}, the problem of defining the electromagnetic momentum in a refractive medium has persisted. However, it seems reasonable that the Minkowski momentum should be treated as a pseudo-momentum, partly depending on material contributions, while the Abraham momentum is a proper electromagnetic momentum (for a discussion, see \cite{Feigel}).

Here we will analyse the radiative transfer equations and derive macroscopic conservation equations in terms of both Abraham's and Minkowski's definitions. It will also be shown that variables can be used as to remove the problem of momentum definition when deriving conservation equations. 
%A short discussion of the refractive influence on acoustic wave propagation is given.
Furthermore, a completely general energy-momentum conservation equation for a radiation fluid in curved spacetime with a refractive medium is derived and discussed. It is shown that the combination of curvature and refraction gives important contributions to the conservation equation, even if the refractive index is spacetime homogeneous. The equation is discussed in the context of cosmological models.

%%%%%%%%%%%%%%%%%%%%%%%%%%%
\section{Ray dynamics and kinetic theory}
%%%%%%%%%%%%%%%%%%%%%%%%%%%

In mechanics, given the action $S$ of a system, we may define the momentum
and Hamiltonian according to
\begin{equation}\label{eq:mechanics}
  p_{\alpha} = \frac{\partial S}{\partial x^{\alpha}} \, \qquad \, 
  H = -\frac{\partial S}{\partial t} ,
\end{equation}
respectively. From this, the equations of motion for a single particle can be 
written in terms of Hamilton's equations, i.e.\
\numparts
\begin{eqnarray}
  && \dot{x}^{\alpha} = \frac{\partial H}{\partial p_{\alpha}} , \\
  && \dot{p}_{\alpha} = -\frac{\partial H}{\partial x^{\alpha}} .
\end{eqnarray}
\label{eq:hamilton}
\endnumparts

In geometric optics \cite{Landau-Lifshitz}, we introduce the eikonal 
$\phi = \phi(t, x^{\alpha})$ by writing the electromagnetic field $F$ 
in the form 
\begin{equation}
  F = a\exp(i\phi) .
\end{equation}
From the eikonal, we define the the wave-vector and frequency of the field as
\begin{equation}
  k_{\alpha} = \frac{\partial\phi}{\partial x^{\alpha}} \, \qquad \, 
  \omega = -\frac{\partial\phi}{\partial t} ,
\end{equation}
in analogy with  (\ref{eq:mechanics}). The equations of ray optics 
becomes
\numparts
\begin{eqnarray}
  && \dot{x}^{\alpha} = \frac{\partial\omega}{\partial k_{\alpha}}
                              , 
                              \label{eq:group1} \\
  && \dot{k}_{\alpha} = -\frac{\partial\omega}{\partial x^{\alpha}} 
                              , 
                              \label{eq:force1}
\end{eqnarray}
\label{eq:ray}
\endnumparts
and from this also follows that $d\omega/dt = \partial\omega/\partial t$. 
These equations are the mass-less quasi-particle analogous of  
(\ref{eq:hamilton}). It is also clear from the eikonal definitions that the 
natural variables describing the photons are the frequency and the 
wave-number rather than the energy and the momentum.

Photons moving in a isotropic non-dispersive but refractive medium will 
satisfy a dispersion relation of the form
\begin{equation}\label{eq:dispersion}
  \omega = \frac{ck}{n}
\end{equation}
where, in general, the refractive index $n$ is spacetime inhomogeneous. 
The equations of motion (\ref{eq:ray}) for individual photons in such a 
medium then becomes
\numparts
\label{eq:ray2}
\begin{eqnarray}
  && \dot{x}^{\alpha} 
       = \frac{c}{n}\hat{k}^{\alpha}, 
       \label{eq:group} \\
  && \dot{k}_{\alpha} 
       = \frac{\omega}{n}\frac{\partial n}{\partial x^{\alpha}} .
       \label{eq:force}
\end{eqnarray}
\endnumparts

Before proceeding further, a discussion of the definition of the 
photon momentum is in place.  
We note that we have chosen as our fundamental variables 
$t, x^{\alpha}$ and $\omega, k_{\alpha}$. The relation to Hamiltonian 
particle dynamics becomes obvious if we choose $H = \hbar\omega$ 
and $p_{\alpha} = \hbar k_{\alpha}$, and it is therefore tempting to assume 
this direct relationship.
However, in electromagnetic theory, there are two distinct ways to define the momentum of the electromagnetic field. According to Minkowski, the momentum is proportional to $\mathbf{D}\times\mathbf{B}$, while according to Abraham it is proportional to $\mathbf{E}\times\mathbf{H}$ (see Ref.\ \cite{Jackson} for a discussion). Furthermore, in measurement of photon momenta in isotropic diffractive media, there are two distinct forms as well, namely $p = \hbar k = \hbar\omega n/c$ and $p \equiv \hbar\omega/nc = \hbar k/n^2$.
We note that the former is in agreement with Minkowski's momentum density and a number of experiments, while the latter definition is consistent with Abraham's choice in terms of the 
Poynting flux (see \cite{Loudon} and references therein), which also follows from detailed microscopic considerations as well as symmetry arguments. 
On the other hand, measurements indicates that $p \propto n$, consistent with 
Minkowski's momentum definition in a dielectric medium. The discrepancy 
between the two definitions can be attributed to the contributions of the medium
in Minkowski's definition, while Abraham's definition (used here) is the `proper' 
momentum of the photon \cite{Feigel}. However, we note that while the form 
(\ref{eq:group}) of the group velocity is valid independently of the momentum 
definition (as expected),  (\ref{eq:force}) formulated in terms of the momentum 
takes different forms for the Minkowski and the Abraham definition.

Given a spectral distribution $\mathscr{N}(t,x^{\alpha},k_{\alpha})$ of photons, the absence of collisions defines the on-shell Vlasov equation for $\mathscr{N}$ according to
\begin{equation}
  \dot{\mathscr{N}} = \frac{\partial\mathscr{N}}{\partial t} 
  + \dot{x}^{\alpha}\frac{\partial\mathscr{N}}{\partial x^{\alpha}} 
  + \dot{k}_{\alpha}\frac{\partial\mathscr{N}}{\partial k_{\alpha}} = 0 ,
  \label{eq:vlasov}
\end{equation}
expressing the phase space conservation of quasi-particles. 

Using the spectral distribution function defined on-shell, i.e.\ with  (\ref{eq:dispersion}) satisfied, we can now define the macroscopic observables, from which a fluid theory can be constructed. 

%%%%%%%%%%%%%%%%%%%%%%%%%%%
\section{Macroscopic variables}
%%%%%%%%%%%%%%%%%%%%%%%%%%%

Next, we define macroscopic variables as moments of the 
distribution function $\mathscr{N}$. In general, we define 
$\langle\psi\rangle = (\int\mathscr{N}d^3k)^{-1} 
\int \psi\mathscr{N} d^3k$ to be the statistical average 
of the function $\psi$, which may be defined over the 
full phase space.  Moreover, in these definitions we closely 
follow Ref.\ \cite{Mihalas}.

--- 
The \emph{number density} $N$ is defined according to
\begin{equation}
  N = \int \mathscr{N}d^3k .
\end{equation}

--- 
The \emph{energy density} $\mu$ is defined as
\begin{equation}
  \mu = \int \hbar\omega\mathscr{N} d^3k .
\end{equation}

--- 
The \emph{average fluid velocity} $u^{\alpha}$ is given by
\begin{equation}
  u^{\alpha} = \langle \dot{x}^{\alpha}\rangle ,
\end{equation}
and from this we may also define the \emph{thermal (or random) 
fluid velocity} $w^{\alpha}$ using
\begin{equation}
  w^{\alpha} = \dot{x}^{\alpha} - u^{\alpha} ,
\end{equation}
such that $\langle w^{\alpha}\rangle = 0$.

--- 
The \emph{radiation energy flux} $q^{\alpha}$ is defined such
 that $q^{\alpha}\,dS_{\alpha}$ is the rate of energy flow across 
 the surface $dS_{\alpha}$. Thus
\begin{equation}
  q^{\alpha} = \int\hbar\omega w^{\alpha}\mathscr{N} d^3k .
\end{equation}
%\end{itemize}
%
%\begin{itemize}

--- 
The \emph{momentum density} ${\mom}^{\alpha}$ of the photon 
in a refractive medium can be formulated in two ways, in accordance 
with Minkowski or Abraham, and they read
\numparts
\begin{equation}\label{eq:momentum-minkowski}
  {\mom}^{\alpha}_M = \int \hbar k^{\alpha}\mathscr{N} d^3k 
  = \int \frac{n^2}{c^2}\hbar\omega \dot{x}^{\alpha}\mathscr{N} d^3k,
\end{equation}
and
\begin{equation}\label{eq:momentum-abraham}
  {\mom}^{\alpha}_A = \int \frac{\hbar k^{\alpha}}{n^2}\mathscr{N} d^3k 
  = \int \frac{1}{c^2}\hbar\omega \dot{x}^{\alpha}\mathscr{N} d^3k ,
\end{equation}
\endnumparts
respectively, where we have made use of  (\ref{eq:dispersion})--(\ref{eq:force}).

--- 
The \emph{pressure tensor} $\mathsf{P}^{\alpha\beta}$ can 
similarly be defined in two distinct ways. Since the pressure 
tensor $\mathsf{P}^{\alpha\beta}$ is defined to be, for an 
observer comoving with the average fluid flow, the rate of 
transport of the $\alpha$ component of momentum per unit 
area of a surface orthogonal to the $\beta$ coordinate (or 
tetrad) axis, we have
\numparts
\begin{equation}\label{eq:pressure-minkowski}
  \mathsf{P}^{\alpha\beta}_M = \int \hbar k^{\alpha} w^{\beta} 
  \mathscr{N} d^3k =  \int \frac{n^2}{c^2} \hbar\omega w^{\alpha} w^{\beta} 
  \mathscr{N} d^3k,
\end{equation}
and
\begin{equation}\label{eq:pressure-abraham}
    \mathsf{P}^{\alpha\beta}_A 
    = \int \frac{1}{n^2}\hbar k^{\alpha} w^{\beta} \mathscr{N} d^3k 
    = \int \frac{1}{c^2}\hbar\omega w^{\alpha} w^{\beta} \mathscr{N} d^3k ,
\end{equation}
\endnumparts
for the Minkowski and the Abraham momentum, respectively. Here we 
have again used  (\ref{eq:dispersion})--(\ref{eq:force}), and we note 
that the pressure tensor is symmetric.

Furthermore, the energy flux and the momentum density is
related to each other through the relations
\numparts
\begin{equation}
  {\mom}^{\alpha}_M = \frac{n^2}{c^2}\left( 
  \mu u^{\alpha} + q^{\alpha} \right) ,
  \label{eq:momentum-energy1}
\end{equation}
and
\begin{equation}
  {\mom}^{\alpha}_A = \frac{1}{c^2}\left( 
  \mu u^{\alpha} + q^{\alpha} \right) ,
  \label{eq:momentum-energy2}
\end{equation}
\label{eq:momentum-energy}
\endnumparts
where we have used general relation $k^{\alpha} = n^2\omega 
(u^{\alpha} + w^{\alpha})/c^2$.

Higher order moments can be defined analogously, 
but the above will suffice for our purposes, and with these 
definitions we are ready to set up the fluid equations. We note 
that the macroscopic quantities defined above are valid also for 
dispersive media, i.e.\ when the refractive index depend on the 
frequency $\omega$ or, in the anisotropic case, the wave vector 
$k^{\alpha}$.

%%%%%%%%%%%%%%%%%%%%%%%%%%%
\section{Fluid equations}
%%%%%%%%%%%%%%%%%%%%%%%%%%%

A hierarchy of fluid equations may be obtained from  
(\ref{eq:vlasov}) by taking the moments with respect to suitable 
microscopic quantities. Given a microscopic variable ${{\psi}}$ 
(which may be a tensorial object), the general fluid conservation 
equation takes the form
\begin{eqnarray}
  \frac{\partial}{\partial t}(N\langle {{\psi}}\rangle) +
  \frac{\partial}{\partial x^{\alpha}}\left( N\langle\dot{x}^{\alpha}{{\psi}}\rangle\right)
  = N\langle\dot{{{\psi}}}\rangle 
  \label{eq:moment}
\end{eqnarray}
where $\dot{{{\psi}}}$ is defined in accordance with  (\ref{eq:vlasov}). 
Thus, $N\langle{{\psi}}\rangle$ and $N\langle\dot{x}^a{{\psi}}\rangle$ 
represent the macroscopic density and macroscopic current, 
respectively, of the microscopic variable ${{\psi}}$, while the right hand 
side of  (\ref{eq:moment}) acts as a source/sink. 
Finally, the fluid hierarchy of equation can be terminated by assuming a 
set of thermodynamic relationships between the macroscopic variables, 
such as an equation of state.

Putting $\psi = 1$, we obtain the conservation equation for the
number of quasi-particles
\begin{equation}
  \frac{\partial N}{\partial t} + \frac{\partial}{\partial x^{\beta}}\left( 
  Nu^{\beta} \right) = 0.
\end{equation}
  
The energy conservation equation is obtained by taking ${{\psi}} = 
\hbar\omega$. Thus, for an observer comoving with the average radiation 
fluid flow we obtain
\begin{equation}\label{eq:energy}
  \frac{\partial{\mu}}{\partial t} + \frac{\partial}{\partial 
  x^{\alpha}}\left( \mu u^{\alpha} + q^{\alpha} \right) =
  -\frac{{\mu}}{n}\frac{\partial n}{\partial t} .
\end{equation}

Similarly, the momentum conservation equation is obtained by 
$\psi = p_{\alpha}$, and takes the form
\numparts
\begin{equation}
  \frac{\partial{\mom}_{M\alpha}}{\partial t} 
  + \frac{\partial}{\partial x^{\beta}}\left( 
  u^{\beta}{\mom}_{M\alpha} + \frac{n^2}{c^2}u_{\alpha}q^{\beta} 
  + \mathsf{P}^{\beta}_{M\alpha}  \right) 
  = \frac{\mu}{n}\frac{\partial n}{\partial x^{\alpha}}
  \label{eq:momentumconservation-M} 
\end{equation}
or
\begin{eqnarray}
  \fl  \frac{\partial{\mom}_{A\alpha}}{\partial t} 
  + \frac{\partial}{\partial x^{\beta}}\left( 
  u^{\beta}{\mom}_{A\alpha} + \frac{1}{c^2}u_{\alpha}q^{\beta} 
  + \mathsf{P}^{\beta}_{A\alpha}  \right)
%   \nonumber \\ &&\quad\quad 
   &=& \frac{\mu}{n^3}\frac{\partial n}{\partial x^{\alpha}} 
  - 2\frac{{\mom}_{A\alpha}}{n}\left( \frac{\partial}{\partial t} 
  + u^{\beta}\frac{\partial}{\partial x^{\beta}} \right)n 
  \nonumber \\
  && 
  - \frac{2}{n}\left( \frac{1}{c^2}u_{\alpha}q^{\beta} 
     + \mathsf{P}^{\beta}_{A\alpha} \right)\frac{\partial n}{\partial x^{\beta}} ,
  \label{eq:momentumconservation-A}
\end{eqnarray}
 \label{eq:momentumconservation1}
\endnumparts
depending on if we use (\ref{eq:momentum-minkowski}) 
and (\ref{eq:pressure-minkowski}) or (\ref{eq:momentum-abraham}) 
and (\ref{eq:pressure-abraham}), respectively. We furthermore note that 
 (\ref{eq:momentumconservation-M}) and 
(\ref{eq:momentumconservation-A}) can be written in a slightly more 
symmetric form, using the relations 
(\ref{eq:momentum-energy1})--(\ref{eq:momentum-energy2}), according to 
\numparts
\begin{equation}
  \frac{\partial{\mom}^{\alpha}_M}{\partial t} 
  + \frac{\partial}{\partial x^{\beta}}\!\!\left( 
  \frac{n^2}{c^2}\mu u^{\alpha}u^{\beta} 
  + \frac{2n^2}{c^2}u^{(\alpha}q^{\beta)} 
  + \mathsf{P}^{\alpha\beta}_{M}  \right) 
  = \frac{\mu}{n}\delta^{\alpha\beta}\frac{\partial n}{\partial x^{\beta}}
\end{equation}
and
\begin{eqnarray}
  \fl \frac{\partial{\mom}^{\alpha}_A}{\partial t} 
  + \frac{\partial}{\partial x^{\beta}}\!\left( 
  \frac{1}{c^2}\mu u^{\alpha}u^{\beta} 
  + \frac{2}{c^2}u^{(\alpha}q^{\beta)} 
  + \mathsf{P}^{\alpha\beta}_A  \right) 
%  \nonumber \\ &&\quad
    &=& \frac{\mu}{n^3}\delta^{\alpha\beta}\frac{\partial n}{\partial x^{\beta}} 
  - \frac{2}{n}{\mom}^{\alpha}_A \frac{\partial n}{\partial t}
  \nonumber \\ && \fl
  - \frac{2}{n}\left( \frac{1}{c^2}\mu u^{\alpha}u^{\beta} 
  + \frac{2}{c^2}u^{(\alpha}q^{\beta)} 
  + \mathsf{P}^{\alpha\beta}_A \right)\frac{\partial n}{\partial x^{\beta}}
\label{eq:momentumcons-A}
\end{eqnarray}
 \label{eq:momentumconservation2}
\endnumparts
respectively. It is straighforward 
to see that  (\ref{eq:momentumconservation-M})--(\ref{eq:momentumconservation-A}) 
can be obtained from each other using the relation ${\mom}^{\alpha}_M 
= n^2{\mom}^{\alpha}_A$.
 
The system of equations presented above can be closed if we choose a 
thermodynamic relationship between certain quantities. 
For an close-to-equilibrium system, the pressure tensor becomes nearly isotropic, 
and we can write $\mathsf{P}^{\alpha\beta} \approx Ph^{\alpha\beta}$, where 
$P = h_{\alpha\beta}\mathsf{P}^{\alpha\beta}/3$. From  (\ref{eq:pressure-minkowski})
and (\ref{eq:pressure-abraham}), we obtain
\begin{equation}\label{eq:pressuredefs}
  P_{M} = \tfrac{1}{3}\mu \, \qquad \, P_{A} = \tfrac{1}{3}n^{-2}\mu ,
\end{equation}
respectively.

The remaining freedom in the equations is removed by choosing 
the observer. We will here use the particle frame, where $u^{\alpha} = 0$.

Using these choices, we obtain a set of equations independent of the choice of
momentum definition, in terms of the energy density and heat flux, according to 
\begin{equation}
  \frac{\partial\mu}{\partial t} + \frac{\partial q^{\alpha}}{\partial x^{\alpha}} 
  = -\frac{\mu}{n}\frac{\partial n}{\partial t} , 
  \label{eq:energy-comoving}
\end{equation}
and
\begin{equation}
   \frac{\partial q^{\alpha}}{\partial t} 
  + \frac{c^2}{3n^2}\delta^{\alpha\beta}\frac{\partial\mu}{\partial x^{\beta}} 
  = \frac{c^2\mu}{n^3}h^{\alpha\beta}\frac{\partial n}{\partial x^{\beta}}
  - \frac{q^{\alpha}}{n}\frac{\partial n}{\partial t} , 
  \label{eq:momentum-comoving}
\end{equation}
from (\ref{eq:momentum-energy1})-- (\ref{eq:momentum-energy2}), (\ref{eq:energy}), and 
(\ref{eq:momentumconservation-M})--(\ref{eq:momentumconservation-A}), 
respectively. We may from this derive 
general wave equations for the energy density and energy flux for radiation
in diffractive media. 
%Next, we investigate the acoustic properties of these equations.

%%%%%%%%%%%%%%%%%%%%%%%%%%%
\section{Covariant conservation laws}
%%%%%%%%%%%%%%%%%%%%%%%%%%%

In this section we will make use of the $1+3$ orthonormal frame (ONF) 
approach (see \cite{Ellis-vanElst} for an overview), since it allows for a 
general relativistic treatment, simplifies calculations, and gives the 
natural Cartesian like reference frame for an observer moving with the 
timelike frame direction. All vector and tensor quantities will be projected 
onto this frame. 

We define a set of frame vectors $\{ \bm{e}_a \}$, $a = 0, \ldots, 3$ 
such that $\bm{e}_a\cdot\bm{e}_b = \eta_{ab}$ gives the constant 
coefficients of the metric: $g_{ab} = \eta_{ab} = \mathrm{diag}\,(-1, 1, 1, 1)$, 
i.e., a Lorentz frame.
The Ricci rotation coefficients $\Gamma^a\!_{bc}$, antisymmetric in the 
first two indices, giving the kinematics of spacetime are defined by
$\nabla_c\bm{e}_b = \Gamma^a\!_{bc}\bm{e}_a$. Here $\nabla_c$ is
the covariant derivative with respect to the Lorentz frame.

The distribution function $\tilde{\mathscr{N}}$ will now be a function 
of the canonical phase space variables $x^a$ and $k_a$, and the 
conservation of phase space density can be written
\begin{equation}\label{eq:fourkinetic}
   \frac{d\tilde{\mathscr{N}}}{d\lambda} = \hat{L}[\tilde{\mathscr{N}}]  
   = \dot{x}^a\bm{e}_a(\tilde{\mathscr{N}}) 
   + \dot{k}_a\frac{\partial\tilde{\mathscr{N}}}{\partial k_a} = \mathscr{C} ,
\end{equation}
where the over dot stands for $d/d\lambda$, where $\lambda$ is an affine parameter along the photon path, and we have introduced the Liouville operator $\hat{L} = \dot{x}^a\bm{e}_a + \dot{k}_a\partial/\partial k_a$. Furthermore, for the sake of generality, we have added the collisional operator $\mathscr{C}$, representing photon emission and absorption \footnote{The process of emission and absorption is thoroughly treated in any book on radiation transport, see e.g.\ \cite{Mihalas}, and will therefore not be discussed further here, apart from the following brief comment. As an example, in case of applications to the early universe, the major contribution to the collisional term would be in the form of Thomson scattering, but there are of course numerous other scattering events that could dominate in other parameter regimes.}. If $f^a$ are the external forces acting on the pencil of light, the covariant derivative along the photon path is  
\begin{equation}
  \frac{Dk_b}{d\lambda} = \dot{k}_b - \Gamma^c\!_{ba}k_c\dot{x}^a = f_b
\end{equation}
With the Hamiltonian $H = H(x^a,k_a)$ in eight dimensional phase space, 
the equations of motion thus reads 
\numparts
\begin{eqnarray}
  && \dot{x}^a = \frac{\partial H}{\partial k_a} , 
  \label{eq:group2} \\
  && \dot{k}_a = -\bm{e}_a(H) + \Gamma^c\!_{ab}k_c\dot{x}^b, 
  \label{eq:force2}
\end{eqnarray}
\label{eq:fourhamilton}
\endnumparts
generalising (\ref{eq:group1})--(\ref{eq:force1}). Here the last term of (\ref{eq:force2}) gives the 
gravitational contribution to the equations of motion. We denote the observer four-velocity by $U^a$, normalised such that $U^aU_a = -c^2$. We partially fix the frame by letting 
the observer four-velocity $U^a = \delta^a_0$, and split spacetime quantities 
with respect to $U^a$. The space metric orthogonal to $U^a$ takes the form
\begin{equation}
  h_{ab} = \eta_{ab} + U_aU_b/c^2 , 
\end{equation}
and the wavevector can be written
\begin{equation}
  k_a = \omega U_a/c^2 + k\ell_a ,
\end{equation}
where $\omega = -U^ak_a$, $k = (h^{ab}k_ak_b)^{1/2}$, and 
$\ell_a = h_a\!^bk_b/k$. We note that $\ell^a\ell_a = 1$ and 
$\ell_aU^a = 0$. In these variables, the dispersion relation 
(\ref{eq:dispersion}) retains its form. 

There is a certain arbitrariness in the choice of Hamiltonian. We choose
the Hamiltonian $H$ such that $H = 0$ gives the dispersion relation 
(and $\partial H/\partial k_a \neq 0$ on the dispersion surface), and the 
variables $x^a, \omega, k$, and $\ell_a$ is treated as independent, with 
the vanishing Hamiltonian as a constraint. Suppose the dispersion 
relation takes the form
\begin{equation}
  \omega = W(x^a, k, \ell_a) .
\end{equation}
Then 
\begin{equation}
  H = -\omega + W(x^a, k, \ell_a) 
\end{equation}
satisfies the necessary criteria listed above, giving the equations of motion
\numparts
\begin{eqnarray}
  && \dot{x}^a = U^a + \frac{\partial W}{\partial k_a}, 
  \label{eq:velocity}\\
  && \dot{k}_a = -\bm{e}_a(W) + \Gamma^c\!_{ab}k_c\dot{x}^b ,
\end{eqnarray}
\endnumparts
via (\ref{eq:group2})--(\ref{eq:force2}), and, moreover, $\dot{H} = \bm{e}_0(H) = \bm{e}_0(W)$. 
With  (\ref{eq:dispersion}), $W(x^a,k) = kc/n(x^a)$, and 
\numparts
\begin{eqnarray}
  && \frac{\partial W}{\partial k_a} = \frac{c\ell^a}{n}, 
  \label{eq:wk} \\
  && \bm{e}_a(W) = - \frac{\omega}{n}\bm{e}_a(n) .
  \label{eq:wx}
\end{eqnarray}
\endnumparts
Thus, with the particular dispersion relation (\ref{eq:dispersion}), 
the kinetic equation (\ref{eq:fourkinetic}) becomes
\begin{equation}\label{eq:fourkinetic2}
  \left( U^a + \frac{c\ell^a}{n} \right)\bm{e}_a(\tilde{\mathscr{N}}) 
  + \left[  \frac{\omega}{n}\bm{e}_a(n) + \Gamma^c\!_{ab}k_c\dot{x}^b\right]
  \frac{\partial\tilde{\mathscr{N}}}{\partial k_a} = \mathscr{C} .
\end{equation}

%With  (\ref{eq:fourhamilton}), the kinetic equation (\ref{eq:fourkinetic}) can be written
%\begin{equation}\label{eq:liouville}
%  \hat{L}[\tilde{\mathscr{N}}] 
%  = \frac{\partial}{\partial x^a}\left( \dot{x}^a\tilde{\mathscr{N}} \right)
%  + \frac{\partial}{\partial k_a}\left( \dot{k}_a\tilde{\mathscr{N}} \right) = 0,
%\end{equation}

In the literature, the common definition of the energy-momentum 
tensor is 
\begin{equation}
  T^{ab}_M = \hbar\int k^ak^b\tilde{\mathscr{N}}\,{\meas} ,
\end{equation} 
where ${\meas} = |\mathrm{det}\,g|^{-1/2}(\delta(H)/\omega)\,d^4k$ 
is the invariant volume measure in momentum space. 
This definition is consistent with the Minkowski definition of the 
photon momentum in diffractive media. Furthermore, in non-diffractive 
media, this is a conserved quantity, due to the one-to-one 
correspondence between $\dot{x}^a$ and $k_a$. However, in diffractive media, 
this correspondence is lost, since 
\begin{equation}\label{eq:x-k}
  \dot{x}^a = c^2\frac{k^a}{\omega} + \frac{c}{n}(1 - n^2)\ell^a 
  = \frac{c^2}{\omega}\left[ g^{ab} - \left( 1 - \frac{1}{n^2} \right)h^{ab} \right]k_b
\end{equation}
[see  (\ref{eq:velocity}) and (\ref{eq:wk})]. Thus, in diffractive media, 
the energy-momentum tensor may alternatively be defined according to 
\begin{equation}\label{eq:en-mom}
  T^{ab}_A = \hbar\int\frac{\omega^2}{c^2} \frac{\dot{x}^a}{c}
  \frac{\dot{x}^b}{c} \tilde{\mathscr{N}}\,{\meas} 
  = \frac{\hbar}{c^4}\int\omega \dot{x}^a\dot{x}^b \tilde{\mathscr{N}}\,\delta(H)\,d^4k , 
\end{equation}
in accordance with Abraham's photon momentum definition. 
We note that when $n \rightarrow 1$, we regain the common definition of 
the energy-momentum tensor in non-diffractive media.  Moreover, since
(\ref{eq:x-k}) holds in general, we have the relation
\begin{equation}
  T^{ab}_A = \left[ g^a\!_c - \left( 1 - \frac{1}{n^2} \right)h^a\!_c \right]%
  \left[ g^b\!_d - \left( 1 - \frac{1}{n^2} \right)h^b\!_d \right]T^{cd}_M 
  \label{eq:transformation}
\end{equation}
between the two definitions of the energy-momentum definition.

From any energy-momentum tensor, we may define the fluid 
quantities used in the preceding sections. We thus have the
relativistic energy density $\mu = T_{ab}U^aU^b$ with respect to 
$U^a$, the relativistic momentum density $\Pi^a = -h^{ab}T_{bc}U^c$,
the scalar pressure $P = (c^2T_{ab}h^{ab})/3$, and the full
anisotropic pressure $\mathsf{P}_{ab} = T_{cd}h^c\!_ah^d\!_b$. By 
construction $\Pi^a$ and $\mathsf{P}_{ab}$ are spacelike 
quantities, i.e., orthogonal to $U^a$.
 
The equations of conservation of energy and momentum can be 
obtained by multiplying  (\ref{eq:fourkinetic2}) by 
$\hbar\omega^2\dot{x}^b/c^4$, and integrating over the momentum 
space variables on the dispersion surface. The result is the 
energy-momentum conservation equation
\begin{eqnarray}
&&\fl  \nabla_aT^{ab}_A  
  = 2\left(1 - \frac{1}{n^2} \right)\Gamma^c\!_{ad}\left[ n^2g^{a(b}h_{ce} - h^{a(b}g_{ce} 
   - n^2\left(1 - \frac{1}{n^2}\right)h^{a(b}h_{ce} 
  \right]T^{d)e}_A  
  \nonumber \\ &&\fl\quad
  + \frac{1}{c^2}\left\{ \left[ g^{ab} - \left(1 - \frac{1}{n^2} \right)h^{ab} \right]U_dU_e
    + 2U^aU_dg^b\!_e - 4 h^{(a}\!_dg^{b)}\!_e\right\}T^{de}_A%
    \frac{\nabla_a n}{n}
  + C^b ,
  \label{eq:conservation}
\end{eqnarray}
where $C^b = \int (\hbar\omega^2\dot{x}^b/c^4) \mathscr{C} \, {\meas}$ 
is the collisional contribution, and $\nabla_a$ is the covariant derivative with respect to the Lorentz frame. We note that as $n \rightarrow 1$, all terms
on the right hand side, except $C^b$, vanishes and we obtain the 
standard conservation equation for non-diffractive media. Moreover,
using the transformation (\ref{eq:transformation}) we obtain the
corresponding equation in terms of the Minkowski variables. 
The first term on the right hand side gives the coupling between spacetime
kinematics and the media refractive properties. Thus, when we have either
flat spacetime or non-diffractive media, this term vanishes. The second term
arises from the spatial and/or time-like dependence of the refractive index. 
Equation
(\ref{eq:conservation}) gives the evolution of energy and momentum
of a radiation fluid in diffractive media on an arbitrary curved spacetime.
Thus, (\ref{eq:conservation}) is the proper starting point for analysing 
photon gas dynamics in general relativistic gravity. 
As compared to the case of a Minkowski background spacetime, there is 
a significant alteration to the equation due to the novel coupling between gravity
and the diffractive media, as given by the first term on the right hand side of 
(\ref{eq:conservation}).

%In the case of a stationary weakly refractive media we have $n^2 \approx 1 + \delta$ ($\delta \ll 1$), and we may linearise (\ref{eq:conservation}) to obtain 
%\begin{eqnarray}
%%&& 
%\fl  \nabla_aT^{ab}_A  
%  \approx 2\delta\Gamma^c\!_{ad}\left( g^{a(b}h_{ce} - h^{a(b}g_{ce} 
%  \right)T^{d)e}_A  
%%  \nonumber \\ &&\qquad
%  + \frac{1}{2c^2}\left( g^{ab}U_dU_e
% - 4 h^{(a}\!_dg^{b)}\!_e\right)T^{de}_A\nabla_a \delta ,
%  \label{eq:conservation-approx}
%\end{eqnarray}
%where we have neglected the collision term

As a simple example of curvature effects, we use (\ref{eq:conservation}) to derive the appropriate energy conservation equation in a Friedmann--Lema\^{i}tre--Robertson--Walker spacetime, i.e. spatial homogeneity and isotropy is assumed. The spacetime is characterised by the time dependent scalar quantities $\mu$ (energy density), $P$ (pressure), $\Theta$ (expansion), and $n$ (refractive index) (see \cite{Ellis-vanElst} for details). Thus, (\ref{eq:conservation}) gives 
\begin{equation}\label{eq:cosm}
  \frac{d\mu}{dt} 
  = -\frac{4}{3}\Theta \mu  - \frac{3\mu}{n}\frac{dn}{dt} ,
\end{equation}
where we have used $P_A = \mu/3n^2$ [see (\ref{eq:pressuredefs})], and $d/dt = U^a\nabla_a$. %Treating the refractive index as a function of the energy density, we obtain the generalised nonlinear energy conservation equation  
%\begin{equation}
%  \frac{d\mu}{dt} = -\frac{4}{3}\frac{\Theta\mu}{1 + 3\mu n^{\prime}(\mu)/n(\mu)}
%\end{equation}
%from (\ref{eq:cosm}), where $n^{\prime} = dn/d\mu$. 
Of course, the analysis of the equation requires a specified density dependence of the refractive index, and therefore also a modified equation of state (for an example, see \cite{AmelinoCamelia,Alexander-Brandenberger-Magueijo}), and will not be pursued further here. 

We note the possibility to formulate (\ref{eq:conservation}) in terms of an effective energy-momentum tensor in certain cases, in particular for homogeneous spacetimes \cite{Triginer-Zimdahl-Pavon}. For example, we may define $\mu_{\mathrm{eff}} = n^3\mu$, so that 
(\ref{eq:cosm}) takes the standard form for this new effective density. In general though, this is not a consistent approach \cite{Triginer-Zimdahl-Pavon}, and can apparently only be overcome by introducing an effective geometry (see, e.g., \cite{Anderson-Spiegel}).

\section{Conclusions}

We have discussed the consequences of the different
electromagnetic momentum definitions, due to Minkowski and 
Abraham respectively, in the context of radiation fluid dynamics.
Starting from a kinetic description, a set of fluid equations was
derived and compared for the different definitions of fluid variables.
It was found that by expressing the equations in terms of certain 
variables, they became independent of the choice of momentum 
definition. Finally, from a general relativistic kinetic theory the 
energy-momentum 
conservation equation valid for a radiation gas in a refractive medium
on an arbitrary spacetime, including collisional effects, was 
derived and discussed in a cosmological setting.

\ack This work was supported by the Swedish Research Council 
through the contract No.\ 621-2004-3217.  
The author would like to thank Gert Brodin and 
Chris Clarkson for helpful discussions.

\end{document}